\documentclass[aps,prl,twocolumn,superscriptaddress]{revtex4-1}

\usepackage[english]{babel}
\usepackage{amsmath,amssymb}
\usepackage{amsfonts}
\usepackage{graphicx}
\usepackage{color}
\usepackage{lipsum}
\usepackage{bm}
\usepackage[colorlinks, linkcolor=red, anchorcolor=green, citecolor=blue]{hyperref}

\newcommand{\beq}{\begin{equation}}
\newcommand{\eeq}{\end{equation}}

\begin{document}
\title {Photon assisted braiding of Majorana fermions in a cavity}
\author{Mircea Trif}
\affiliation{Institute for Interdisciplinary Information Sciences, Tsinghua University, Beijing 100084, China}
\author{Pascal Simon}
\affiliation{Laboratoire de Physique des Solides, CNRS, Univ.\ Paris-Sud,
University Paris-Saclay, 91405 Orsay Cedex, France}
\date{\today}

\begin{abstract}
We study the dynamical process of braiding  Majorana bound states  in the presence of the coupling to photons in a microwave cavity.  We show theoretically that the $\pi/4$ phase associated with the braiding of Majoranas, as well as the parity of the ground state are imprinted into the photonic field of the cavity, which can be detected by dispersive readouts techniques. These manifestations  are purely dynamical, they occur in the absence of any splitting of the MBS that are exchanged, and they disappear in the static setups studied previously. Conversely, the cavity can affect the braiding phase, which in turn should  allow for cavity controlled braiding. 
\end{abstract}

\maketitle

 \emph{Introduction.---} Braiding of non-Abelian anyons lies at the heart of topological quantum information processing \cite{SarmaRMP2008}. One promising class of non-abelian anyons are the Majorana bound states (MBS) that emerge in the so called topological  superconductors as  zero-energy quasiparticles~\cite{kitaev2001unpaired,dassarma-majarona-review,aasen2016-qc}. In recent years there have been a great deal of excitement towards detecting and manipulating MBS in various condensed matter platforms~\cite{alicea2012new,Leijnse2012,franz2015review,Aguado2017}. In particular, implementations based on one-dimensional (1D) semiconducting wires (SW) have attracted the most attention. 
Following theoretical proposals~\cite{lutchyn2010majorana,oreg2010helical}, several experiments~\cite{mourik2012signatures,xu2012,Das2012,churchill2013,albrecht2016exponential,Deng2016,nichele2017transport,Kouwenhoven2018} have reported characteristic transport signatures in the form of a zero-bias conductance peak compatible with the presence of zero-energy MBS (see Ref. \cite{Lutchyn2018} for a recent review). 
Nevertheless,  the puzzling question of whether such zero-bias peaks are due to MBS is still under debate~\cite{dassarma2017,tewari2018}  
 and other direct measurable  manifestations of Majorana physics are timely.

The smoking gun feature associated with the MBS is their exchange, or braiding statistics: moving these quasiparticles around each other and exchanging their positions will implement, within the degenerate subspace they pertain to, non-Abelian unitary transformations that depend only on the topology of the trajectories \cite{Kitaev2003,Alicea2011,Sau2011,vanHeck2012}. Such unitary transforms are more robust against decoherence and dephasing due to {\it local} environments, as opposed to quantum computing with conventional qubits 
(although recent works show that such protection might be fragile when going beyond the adiabatic approximation or assuming a coupling of the MBS to  external baths 
\cite{Pedrocchi2015a,Pedrocchi2015b,KnappPRX2016,Bauer2018,SauCM2018}). In order to detect the Majorana signatures, as well as to manipulate the braiding of the MBS, one needs to resort various interference schemes, and eventually lift the ground state degeneracy, thus making MBS interact \cite{alicea2012new,aasen2016-qc,dassarma-majarona-review}. 

In this Letter, we analyze the braiding of Majorana fermions in a tri-junction geometry assisted by photons in a cavity quantum electrodynamics (cQED) setup. We show that both the parity of the ground state and the Berry phase associated with the braiding statistics  imprint into the cavity field,  which in turn can be addressed by conventional dispersive readouts techniques. The present  Berry phase coupling mechanism, which is due to the interplay of dynamics during braiding and the {\it non-locality} of the photonic field, works even when the lowest energy subspace spanned by the Majorana fermions is degenerate at all times. These effects are purely dynamical and do not have any static analogue (e.g. by comparing the beginning and the end of the braiding by cQED spectroscopy). 

\begin{figure}[t] 
\centering
\includegraphics[width=0.99\linewidth]{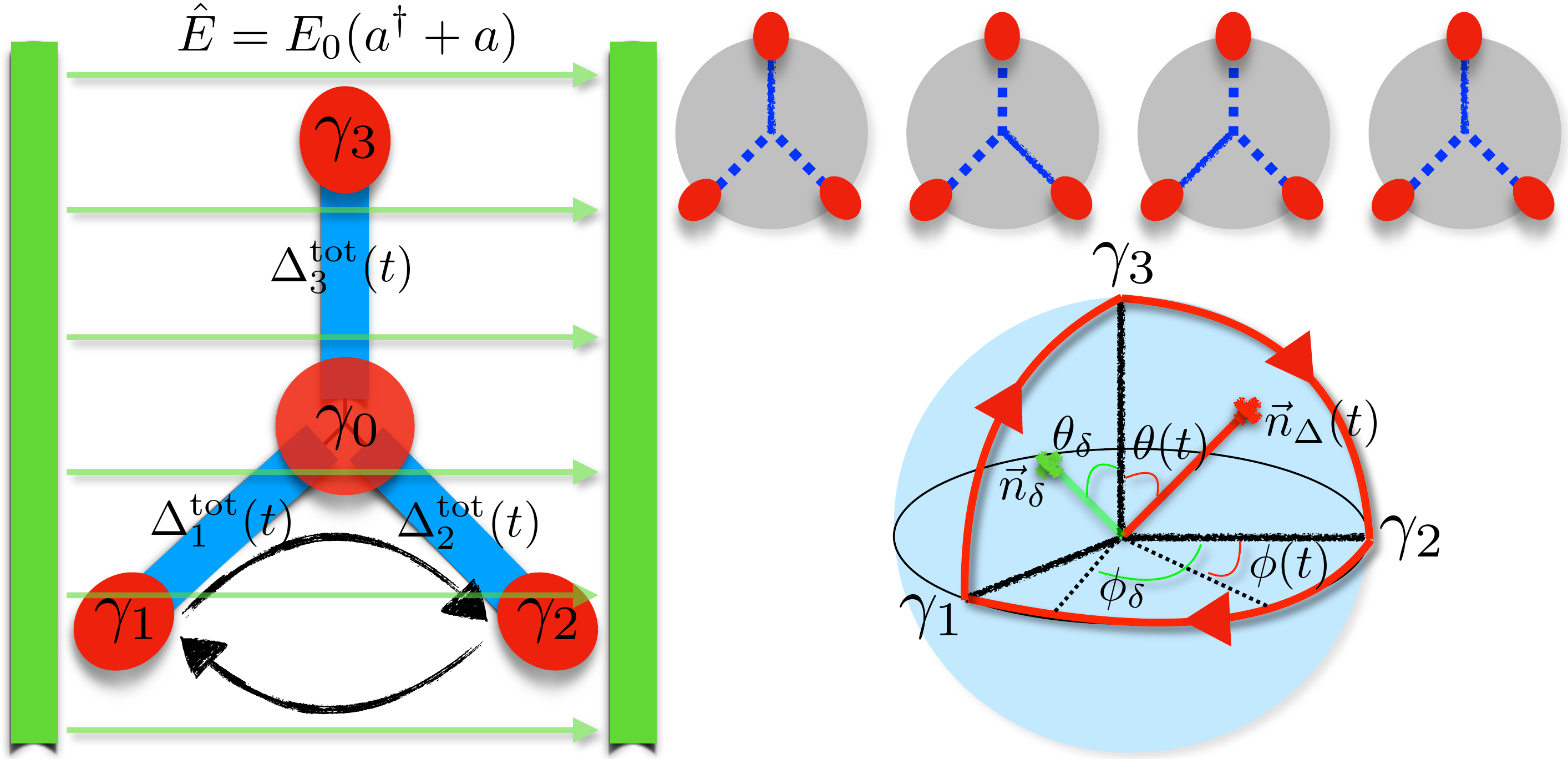}\
\caption{Sketch of the system and the cavity-assisted Majorana braiding (exchange) process of the MBS $\gamma_1$ and $\gamma_2$. Left: The Y junction with the end $\gamma_{i}$, $i=1,2,3$,   and the middle $\gamma_0$ MBS, along with the couplings $\Delta_{i}^{\rm tot}(t)=\Delta_i(t)+\delta_i(a^\dagger+a)$ in the presence of the cavity field $\hat{E}$ (green). Right: the braiding  steps, with the dashed line $i$ corresponding to a splitting $\Delta_i\gg\Delta_{j}$, with $i\neq j$ (top), and the evolution of the vector $\vec{n}_\Delta(t)$ in Eq.~\eqref{Eq} on the Bloch sphere during the braiding process (bottom). The octant spanned by this vector connects to the Berry phase accumulated by the ground state wavefunction that pertains to the Majorana braiding statistics. }
\label{fig1} 
\end{figure}

\emph{System and Model Hamiltonian.---} We consider a Majorana Y junction coupled to the electromagnetic field inside a microwave cavity, as depicted in Fig.~\ref{fig1}. The time-dependent effective Hamiltonian describing the Y junction coupled to the cavity  reads \cite{KarzigPRX2016}:
\begin{align}
H_{\rm Y}(t)&=\frac{i}{2}\gamma_0\vec{\Delta^{\rm tot}}(t)\cdot\vec{\gamma}+\omega_0 a^\dagger a\,, \nonumber\\
\vec{\Delta}^{\rm tot}(t)&=\vec{\Delta}(t)+\vec{\delta}(a^\dagger+a)\,,
\label{Eq}
\end{align}
where $\gamma_0$ stands for the middle MBS (in the center of the junction), $\vec{\gamma}=(\gamma_1,\gamma_2,\gamma_3)$ and $\vec{\Delta}=(\Delta_1,\Delta_2,\Delta_3)$, with $\gamma_i$ and $\Delta_i$ being the outer MBS and the  coupling strengths between  $\gamma_0$ and $\gamma_i$, with $i=1,2,3$. Moreover, $\vec{\delta}=(\delta_1,\delta_2,\delta_3)$ so that $\delta_i=\alpha_i\Delta_i$ for $i=1,2,3$,  with $\alpha_{i}$ the weights of the couplings,  so that  $\delta_i$ vanish when $\Delta_i$ vanishes, as expected for far apart Majoranas \cite{KnappPRX2016,SauCM2018}. Finally,  $a$ ($a^\dagger$) the annihilation (creation) operators for the photons, and $\omega_0$ is the cavity frequency. 

Next we introduce usual fermionic operators in terms of the Majorana ones $c_1=(\gamma_1-i\gamma_2)/2$ and $c_2=(\gamma_0-i\gamma_3)/2$,  which in turn allows us to write the Majorana Hamiltonian only in the basis $\{|00\rangle, c_1^\dagger|00\rangle=|10\rangle,c_2^\dagger|01\rangle=|00\rangle,c_1^\dagger c_2^\dagger|00\rangle=|11\rangle\}$ (see SM\cite{sm} for details):
\begin{align}
H_{\rm Y}^\tau(t)&=\frac{1}{2}\vec{\Delta}_{\tau}^{\rm tot}(t)\cdot\vec{\sigma}+\omega_0a^\dagger a~,
\label{Rabi}
\end{align} 
with $\vec{\Delta}_{\tau}^{\rm tot}(t)=[-\Delta_{2,\tau}^{\rm tot}(t),\tau\Delta_{1,\tau}^{\rm tot}(t),\Delta_{3,\tau}^{\rm tot}(t)]$, and where the Pauli matrices $\sigma_{x,y,z}$ acts within the same parity states $\tau=\pm$, the eigenvalues of the parity operator $\tau_z=\gamma_0\gamma_1\gamma_2\gamma_3$. We see that the pair of states $\{|00\rangle,|11\rangle\}$ and $\{|01\rangle,|10\rangle\}$ do not couple with each other, which is a consequence of the parity  conservation in the system. The Hamiltonian in Eq. \eqref{Rabi} describes the more common spin $1/2$ coupled to a cavity mode (similar to the Rabi model).  

It is instructive to work in spherical coordinates, and  define $\vec{\Delta}_\tau(t)=\Delta(t)\,\vec{n}^\tau_\Delta$, and $\vec{\delta}_\tau=\delta\,\vec{n}_\delta^\tau$, with $\Delta(t)=\sqrt{\sum_{i}\Delta_i^2(t)}$,  $\delta(t)=\sqrt{\sum_{i}\delta_i^2(t)}$, and $\vec{n}^\tau_{\alpha}=(\sin{\theta_\alpha}\cos{\phi_\alpha},\tau\sin{\theta_\alpha}\sin{\phi_\alpha},\cos{\theta_\alpha})$, where $\alpha=\Delta,\delta$ (nevertheless, in the following we drop the $\Delta$ index from the angles of $\vec{n}_\Delta^\tau$). We see that the instantaneous ground state for each of the two parities in the absence of the coupling to the cavity are degenerate, with energy $E_{gs}(t)\equiv-\Delta(t)$. In the following, we assume that all manipulations are adiabatic, or $\dot{\Delta}(t)/\Delta(t)\gg\Delta(t)$ at all times, so that there are no real excitations outside the degenerate subspace and the system always stays in the ground state. 

\emph{Rotating frame description.---} In order to describe the braiding in the adiabatic limit, we perform a time-dependent unitary transformation $A_\tau(t)$ on the Y junction  Hamiltonian, where the columns of the matrix $A_\tau(t)$ are given by the instantaneous eigenstates of $\vec{\Delta}_{\tau}(t)\cdot\vec{\sigma}$. The new transformed Hamiltonian becomes $\tilde{H}_{\rm Y}(t)=A_\tau^\dagger(t)H_{\rm Y}(t)A_\tau(t)-iA^\dagger_\tau(t)\partial_tA_\tau(t)$, or in more details \cite{sm}:
\begin{align}
\!\tilde{H}_{\rm Y}^\tau(t)&=\frac{\Delta(t)}{2}\sigma_z+\tau\frac{\dot{\phi}}{2}(1+\cos{\theta}\sigma_z+\sin{\theta}\sigma_x)\nonumber\\
&\!\!\!\!\!\!+\frac{\dot{\theta}}{2}\sigma_y+A_\tau^\dagger(t)[\vec{\delta}_\tau(t)\cdot\vec{\sigma}]A_\tau(t) (a^\dagger+a)+\omega_0a^\dagger a\,.
\end{align}   
We now proceed with several approximations. First, in the adiabatic limit, we can neglect the terms $\propto\dot{\phi}\sigma_x$ and $\propto\dot{\theta}\sigma_y$  above, as they act in higher order in perturbation theory in $1/\Delta(t)$, while the term $\propto\dot{\phi}\sigma_z$ is diagonal and responsible for the Berry phase contribution to the dynamics. Second, we assume the weak coupling limit, and perform the rotating wave approximation (RWA) which means we keep only the rotating  terms  in the above transformed Hamiltonian. Putting all together, we get the full RWA (time-dependent) Hamiltonian
\begin{align}
\!\!\tilde{H}^\tau_{\rm Y}(t)&\approx \frac{1}{2}\Delta^{\rm eff}_\tau(t) \sigma_z+[\tilde{\delta}_\tau(t)a\sigma_++{\rm h. c. }]+\omega_0a^\dagger a\,,\\
\tilde{\delta}_\tau(t)&=\delta(t)\left[\cos{\theta_\delta}\sin{\theta}-\sin{\theta_\delta}\cos{\theta}\cos{(\phi-\phi_\delta)}\right.\nonumber\\
&\left.+i\tau\sin{\theta_\delta}\sin{(\phi-\phi_\delta)}\right]/2\,,
\end{align}    
where $\Delta^{\rm eff}_\tau(t)=\Delta(t)+\tau\dot{\phi}\cos{\theta}$. We see that in this description, the effective Majorana splitting, $\Delta^{\rm eff}_\tau(t)$, contains a Berry phase contribution with opposite signs for opposite parities $\tau$, and that $\tilde{\delta}_-(t)=\tilde{\delta}^*_+(t)$, which also implies that  $|\tilde{\delta}_+(t)|=|\tilde{\delta}_-(t)|\equiv\tilde{\delta}(t)$ (independent of $\tau$). This Hamiltonian, which is of  Jaynes €"Cummings type,  only couples the states within the pairs $\{|n\uparrow\rangle,|n+1\downarrow\rangle\}$, with $n$ the number of photons in the cavity. Moreover, we can write $\tilde{\delta}_\tau=\tilde{\delta}(t)\exp{[i\Phi_\tau(t)]}$, with $\Phi_\tau(t)=\arg{\tilde{\delta}_\tau(t)}\equiv\tau\Phi(t)$, i.e. the spin directly affects the photonic field by adding a time and parity dependent phase. In the static case ($\dot{\phi}=0$), the eigenspectrum  is the same for the two parities: 
\begin{align}
\epsilon_{n,\tau}=(n+1/2)\omega_0\pm\frac{1}{2}\sqrt{(\Delta-\omega_0)^2+4|\tilde{\delta}|^2(n+1)},
\end{align} 
and thus the cavity cannot discriminate between the two. 
Note that in this limit, some  overlapping between the Majoranas and thus a  splitting of the MBS is required for the cavity to be sensitive to the parity \cite{Dmytruk2015,Dartiailh2016}. Nevertheless, the initial degenerate states built from the MBS become dressed by the photonic field to give rise to degenerate Majorana polaritonic states \cite{Trif2012}. Switching on the dynamics implies manipulating the Majorana polaritonic state, either in the resonant or the dispersive regime. In this work, we only focus on the latter, which allows for detection of the braiding statistics while we leave the resonant case for a future study (in such a case the braiding could be even manipulated by means of the cavity).  

\emph{Dispersive time-dependent Hamiltonian.---} Next we address the case of large detuning, quantified by the conditions $\delta\ll|\Delta(t)-\omega|$. The coupling term $\propto\delta$ is now off-diagonal, and we can treat it in time-dependent perturbation theory. For that, we perform a time-dependent Schrieffer-Wolff unitary transformation $U_\tau(t)=\exp{[S_\tau(t)]}$, with $S_\tau^\dagger(t)=-S_\tau(t)$ on the Hamiltonian $\tilde{H}_{\rm Y}(t)$, chosen as $S_\tau(t)=\tilde{e}_\tau(t)\sigma_+a-\tilde{e}_\tau^*a^\dagger\sigma_-$ \cite{Goldin2000} so that, up to second order in $\delta$, the effective Hamiltonian reads:
\begin{align}
\!\!\!\tilde{H}^\tau_{\rm Y}(t)\approx
\frac{1}{2}\Delta^{\rm eff}_\tau(t)\sigma_z+g_{\tau}(t)\left(a^\dagger a+\frac{1}{2}\right)\sigma_z+\omega_0a^\dagger a\,,
\end{align}
with $g_\tau(t)={\mathcal Re}[\tilde{\delta}_\tau(t)\tilde{e}^*_\tau(t)]$, and where $\tilde{e}_\tau(t)$ is found from the equation
\begin{align}
&[\omega_0-\Delta^{\rm eff}_\tau(t)]\tilde{e}_\tau(t)+\tilde{\delta}_{\tau}(t)+i\dot{\tilde{e}}_\tau(t)=0. 
\label{etau}
\end{align}
In the following, we focus  only on cyclic (periodic) trajectories, so that $\tilde{H}^\tau_{\rm Y}(t+T)=\tilde{H}^\tau_{\rm Y}(t)$, with $T=2\pi/\Omega$ the period of the drive (and $\Omega$ the corresponding frequency). In such a case, the solution to the  equation \eqref{etau} is found as
\begin{align}
\tilde{e}_\tau(t)=\sum_n\frac{\tilde{\delta}_{\tau,n}e^{in\Omega t}}{\Delta_{0}-\omega_0+(n+\tau\gamma_B/2\pi)\Omega}\,,
\end{align}
where $\tilde{\delta}_{\tau,n}=(1/T)\int_0^T dt\tilde{\delta}_\tau(t)\exp{(-2i\pi nt/T)}$ are the Fourier components of $\tilde{\delta}_{\tau}(t)$, $\Delta_{0}=(1/T)\int_0^T dt\Delta(t)$ is the average energy of the effective spin over one period $T$, while $\gamma_B\equiv\gamma_{B}^\uparrow-\gamma_{B}^\downarrow$ denotes the difference of the Berry phase associated with the spin up and spin down trajectories during the cycle. The (spin-dependent) photonic evolution operator during one cycle becomes $U_{\rm eff}(T)=\exp{[-ig^\tau_{\gamma_B}Ta^\dagger a\sigma_z]}$, with  $g^\tau_{\gamma_B}=(1/T)\int_0^Tdtg_\tau(t)$ \cite{sm} or:
\begin{align}
\!g^\tau_{\gamma_B}(\omega_0,\Omega)=\sum_n\frac{|\tilde{\delta}_{r,n}|^2+|\tilde{\delta}_{i,n}|^2+2\tau{\mathcal Im}[\tilde{\delta}_{r,n}\tilde{\delta}^*_{i,n}]}{\Delta_{0}-\omega_0+(n+\tau\gamma_B/2\pi)\Omega}\,,
\end{align}  
where $\tilde{\delta}_{r(i),n}$ are the $n$-th Fourier components of the real (imaginary) parts of $\tilde{\delta}_\tau(t)$. 
This is one of our main results: the  ground state ($\sigma_z=-1$) parity and the associated Berry phases imprints into the photonic field evolution operator during one braiding period $T$. The effect of the parity is two-fold: it affects the matrix elements $\tilde{\delta}_{n}$ in the numerator, and it enters in the denominator via the Berry phase. While the denominator has a very simple (universal) form in terms of the parity via the Berry phase, that seems not to be the case for the numerator. To put these two contributions on equal footing,  we note that generally  $|\tilde{\delta}(t+T)|=|\tilde{\delta}(t)|$, and $\Phi(t+T)=2\pi k+\Phi(t)$, with $k\in {\mathcal Z}$ being the number of times the phase $\Phi(t)$ winds during one period $T$. Writing $\tilde{\delta}_{\tau}(t)\approx \delta_{0}\exp{(ik\tau\Omega t)}$, with $\delta_{0}=(1/T)\int_0^Td\tau|\tilde{\delta}(t)|$, we obtain:
\begin{equation}
g_{\gamma_B}^\tau(\omega_0,\Omega)\approx\frac{\delta_0^2}{\Delta_{0}-\omega_0+\tau(k+\gamma_B/2\pi)\Omega}\,,
\end{equation} 
from which we can now simply read the two parity dependent effects: one contribution from the Berry phase of the effective spin, and another contribution from the winding number $k$ of the phase $\Phi(t)$.  The static case, corresponding to $\gamma_B=2\pi$ and $k=-1$, results in no difference between the two parities, as expected. Note that close to resonances, the  perturbative calculation presented above breaks down. However, as shown later, in the presence of dissipation the parity-dependent resonances can be probed. 
\begin{figure}[t] 
\centering
\includegraphics[width=0.99\linewidth]{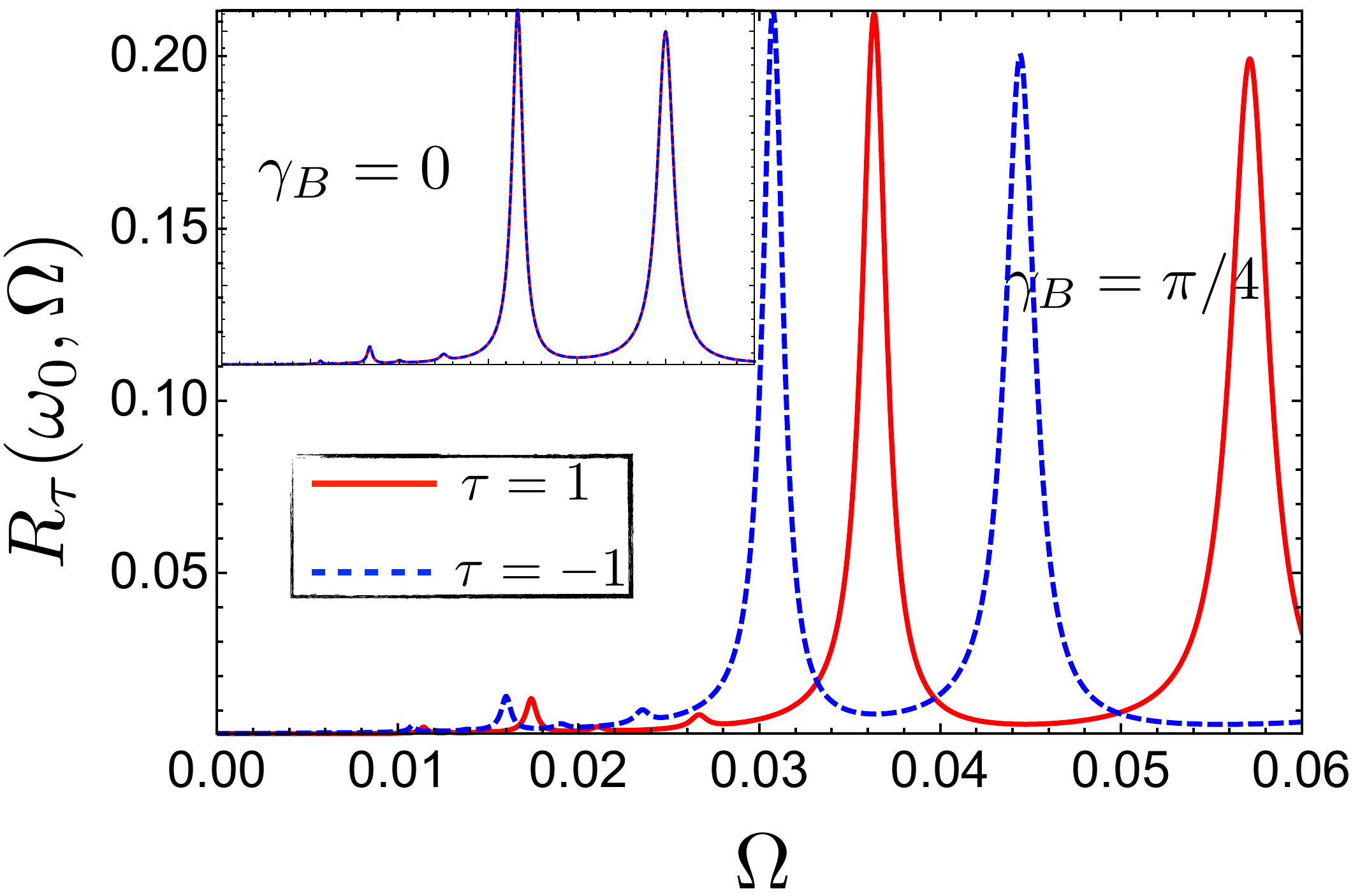}\
\caption{Dependence of the reflection coefficient $R_\tau(\omega_0,\Omega)$ with respect to the braiding frequency $\Omega$ for the two parities $\tau=\pm1$.  In the main (inset) plot we show $R_\tau(\Omega)$ with the Berry phase contribution $\gamma_B=\pi/4$  included (neglected) in the reflection coefficient. The red (full) and the blue (dashed) correspond to parity $\tau=1$ and $\tau=-1$, respectively. The position of the peaks shift according to Eq.~\eqref{shift}, for the braiding trajectory  $\gamma_B=\pi/4$ (main), while they coincide at $\gamma_B=0$ (inset). We consider $\omega=1$, $\Delta_{0}=1.1$,  $\kappa=\Gamma=0.002$, as well as  $\alpha_1=\alpha_2=0$, and $\alpha_3=0.01$, with all energies expressed in terms of the cavity frequency $\omega_0$.}
\label{fig2} 
\end{figure}

\emph{Majorana braiding.---} Here we describe briefly the specific implementation of the braiding (exchanging) of the Majorana fermions $\gamma_1$ and $\gamma_2$. For simplicity, we assume that $\Delta(t)=\Delta_0$ is constant during the entire  dynamics, and consider the following steps showed in Fig.~\ref{fig1}:  ${\bm \Delta}(t)=\Delta_0[0,\sin{\theta(t)},\cos{\theta(t)}]$, for $t\in[0,T/3)$, ${\bm \Delta}(t)=\Delta_0[\sin{\phi(t)},\cos{\phi(t)},0]$, for $t\in[T/3,2T/3)$, and ${\bm \Delta}(t)=\Delta_0[\sin{\theta(t)},0,\cos\theta(t)]$, for $t\in[2T/3,T)$, with $\theta(0)=\theta(T)=\phi(T/3)=0$, $\theta(T/3)=\phi(2T/3)=\pi/2$.  As showed, for example,  in Ref.~\cite{KarzigPRX2016}, this corresponds to the exchange of  the MBS $\gamma_1$ and $\gamma_2$. The specific  implementation for $\theta(t)$ and $\phi(t)$ strongly affects the validity of the adiabatic approximation, especially at the turning points \cite{KnappPRX2016,SauCM2018}.  We disregard such diabatic effects in this work, and for simplicity assume that  $\theta(t),\phi(t)=\Omega t$ on the intervals over which they vary. In the spin language, the Berry phase accumulated by the ground state wavefunction of parity $\tau=\pm$ for a path ${\mathcal C}$ on the sphere in Fig.~\ref{fig1} is calculated from
\begin{equation}
\gamma_{B,\tau}=\frac{1}{2}\int_{{\mathcal C}}d\phi_\Delta[1+\tau\cos{\theta}_\Delta(t)]\,,
\end{equation}
which for the braiding path leads to $\gamma_{B,\tau}=\tau\pi/4$. Using that same prescription allows us to evaluate  $\tilde{\delta}_{\tau,n}$ that enter the photonic evolution operator $U_{\rm eff}(T)$, which will be discussed in the next part. We note that in Ref.~\onlinecite{KarzigPRX2016}  various other paths were studied besides the one pertaining to braiding of MBS,  such as the $\pi/8$ magic gate. Our prescription applies for that case too,  and  such phases could be imprinted  into the photonic field and eventually read, for example, by measuring the output field from the cavity, as discussed in the next part.

\emph{Detection and Dissipation.---} Here we briefly discuss the detection of the cavity field, and consequently the braiding via the the input-output scheme by adding the coupling of the cavity to the external world (or stripline) $H_{c-b}=-i\sum_{k}(f_{k}a^\dagger b_{k}-f_{k}^*b^\dagger_{k}a)$, where $b_{k}$ stand for the external line modes with wavevector $k$. That endows with the following equation for the cavity field:
\begin{align}
\dot{a}=-\left(i\omega_{0}+\frac{\kappa}{2}\right)a-\sum_{j}\sqrt{\kappa}_jb_{in,j}+\tilde{\delta}_\tau(t)\sigma_-\,,
\end{align} 
where $\kappa=\sum_j\kappa_j$,  $\kappa_j\approx\sum_{k}|f_{k,j}|^2\delta(\omega_{k,j}-\omega_0)$ and $b_{in,j}(t)$ are the decay rate of the cavity and the input field onto the cavity used to probe the braiding at port $j=1,2$, respectively. Moreover, the output field $b_{out,j}(t)$  relates to the input and the cavity fields via the relation $b_{in,j}(t)+b_{out,j}(t)=\sqrt{\kappa}_ja(t)$. Switching to the frequency space, we can write $b_{out,1}(\omega)=t_\tau(\omega,\Omega)b_{in,2}(\omega)$ \cite{kohlerPRL2018}, with 
\begin{equation}
t_\tau(\omega,\Omega)\simeq\frac{i\sqrt{\kappa_1\kappa_2}}{\omega_0-\omega-i\kappa/2-g_{\gamma_B}^\tau(\omega+i\Gamma,\Omega)}\,,
\end{equation}
being the complex transmission of the cavity in the presence of the Y junction, where we assumed a finite broadening  $\Gamma$ of the Y Majorana states (see \cite{kohlerPRL2018,sm} for a derivation). 
\begin{figure}[t] 
\centering
\includegraphics[width=0.99\linewidth]{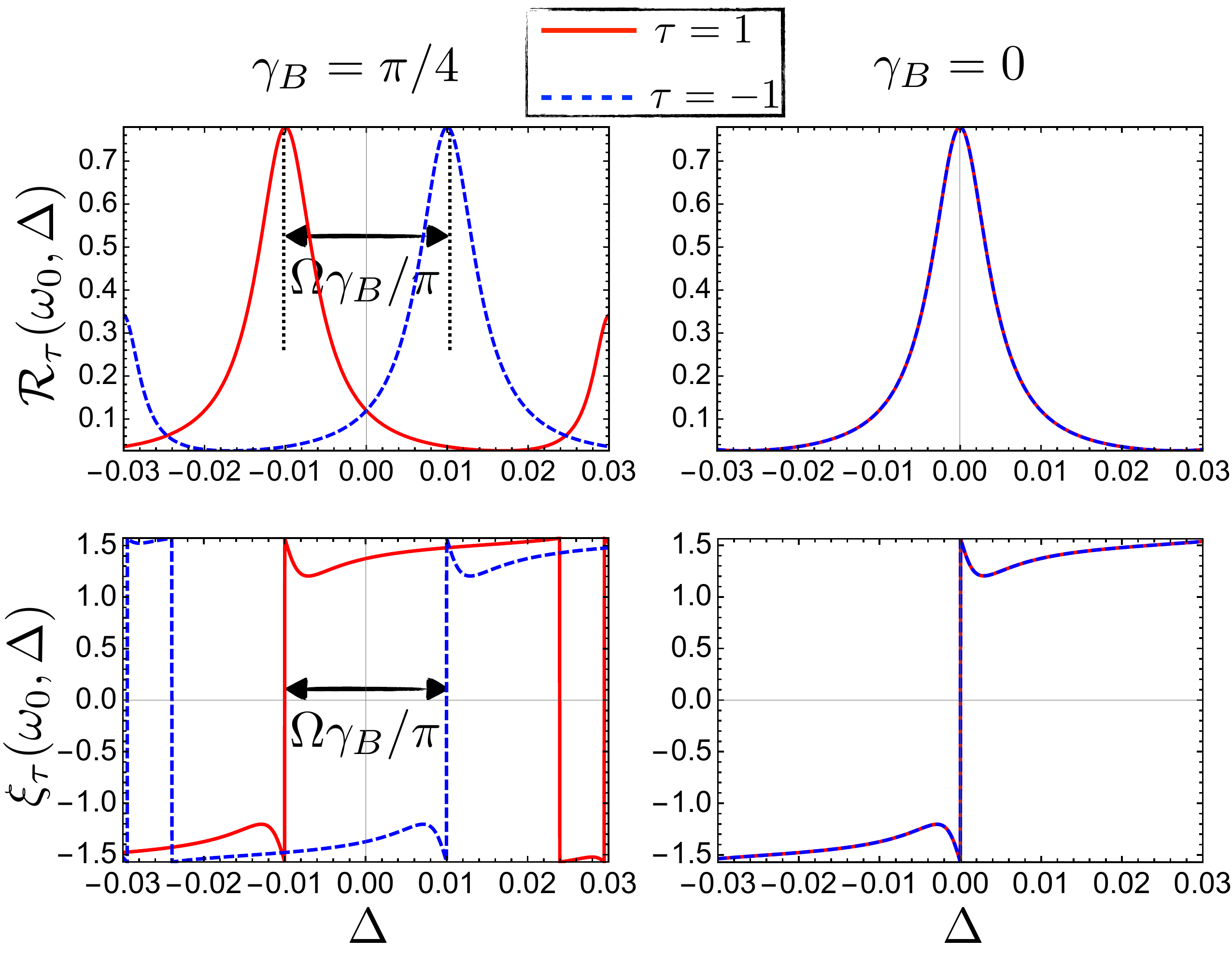}\
\caption{Reflection coefficient $R_\tau(\omega_0,\Delta)$  (upper row)  and the phase shift $\xi_\tau(\omega_0,\Delta)$ (lower row) as a function of the detuning $\Delta=\Delta_{0}-\omega_0$ for the two parities $\tau=\pm1$.  
In the left (right) columns we show the plots for $\gamma_B=\pi/4$  included (neglected), respectively. The red (full) and the blue (dashed) correspond to parity $\tau=1$ and $\tau=-1$, respectively. The position of the peaks shift between the two parities is $\Omega\gamma_B/\pi$. The values of the parameters are the same as in Fig. \ref{fig2}.}
\label{fig3} 
\end{figure}

Typically, both the amplitude $T_\tau(\omega,\Omega)=|t_\tau(\omega,\Omega)|^2$ and the phase $\xi_\tau(\omega,\Omega)=\arctan[{\mathcal Im}(t_\tau)/{\mathcal Re}(t_\tau)]$ of the output signal are measured. In the main (inset) plot of  Fig.~\ref{fig2} we show the reflection coefficient $R_\tau(\omega_0,\Omega)=1-T_\tau(\omega_0,\Omega)$ as a function of the driving frequency $\Omega$ for the two parities when we account for (neglect) the Berry phase contribution pertaining to the braiding of MBS. Moreover, we assume the cavity is probed at resonance $\omega=\omega_0$, and in this case we find the distance between the inverse of $n$-th order peaks of the two parities to satisfy
\begin{equation}
\gamma_B=\pi(\Delta_0-\omega_0)\left(\frac{1}{\Omega_{n,-1}}-\frac{1}{\Omega_{n,+1}}\right)\,,
\label{shift}
\end{equation}
where $\Omega_{n,\tau}$ denotes the $n$-th resonance peak in the braiding frequency $\Omega$, for parity $\tau$. In an experiment, one can therefore extract the Berry phase associated with the braiding by measuring precisely this scaling from the transmission spectrum as a function of $\Omega$ when the cavity assists the braiding. 

Similarly, to illustrate the versatility of our proposal,  in Fig. \ref{fig3} we plot $R_\tau(\omega_0,\Delta)$ and the phase shift $\xi_\tau(\omega_0,\Delta)$ as a function of the detuning parameter  $\Delta$ in the presence (and absence) of the Berry phase $\gamma_B=\pi/4$. One can thus extract the Berry phase from the shift of the resonances for the two different parities from both quantities.
 
\emph{Conclusions.---} 
We have studied the braiding  of Majorana fermions in an Y junction geometry that is embedded in a microwave cavity. We have shown that both the parity of the ground state and the non-trivial Berry phase occurring during the braiding cycle imprint into the photonic field of the cavity that assists the process, which in turn can be probed non-invasively by the dispersive readout of the cavity microwave transmission. We found that these manifestations are purely dynamical, they occur in the absence of any splitting of the MBS that are exchanged, and they disappear in the static case. While here we focus on the effects of braiding on the photons, the reverse effect, namely the manipulation of braiding by means of the photonic field can be analyzed within the same framework.      

\emph{Acknowledgments.---} 
We would like to acknowledge discussions with  Yipu Song, Kihwan Kim, Luyan Sun, and Yaroslav Tserkovnyak.  MT acknowledges support from the National Basic Research Program of China Grants No. 2011CBA00300 and No. 2011CBA00302. 


%

\end{document}


\title {Supplementary Material\\
``Photon assisted braiding of Majorana fermions in a cavity''}

\author{Mircea Trif}
\affiliation{Institute for Interdisciplinary Information Sciences, Tsinghua University, Beijing, China}

\author{Pascal Simon}
\affiliation{Laboratoire de Physique des Solides, CNRS, Univ. Paris-Sud,
Universit\'e Paris Saclay, 91405 Orsay cedex, France}
\date{\today}

\begin{abstract}
In this Supplementary Material (SM) we provide details about the effective Y-junction Hamiltonian, the derivation of the low-energy MBS-photon coupling, as well as details on the the input-output theory for periodically driven systems in the adiabatic limit with dissipation.   
\end{abstract}


\maketitle

\section{Effective four MBS in a Y-junction geometry}

\begin{figure}[t] 
\centering
\includegraphics[width=0.6\linewidth]{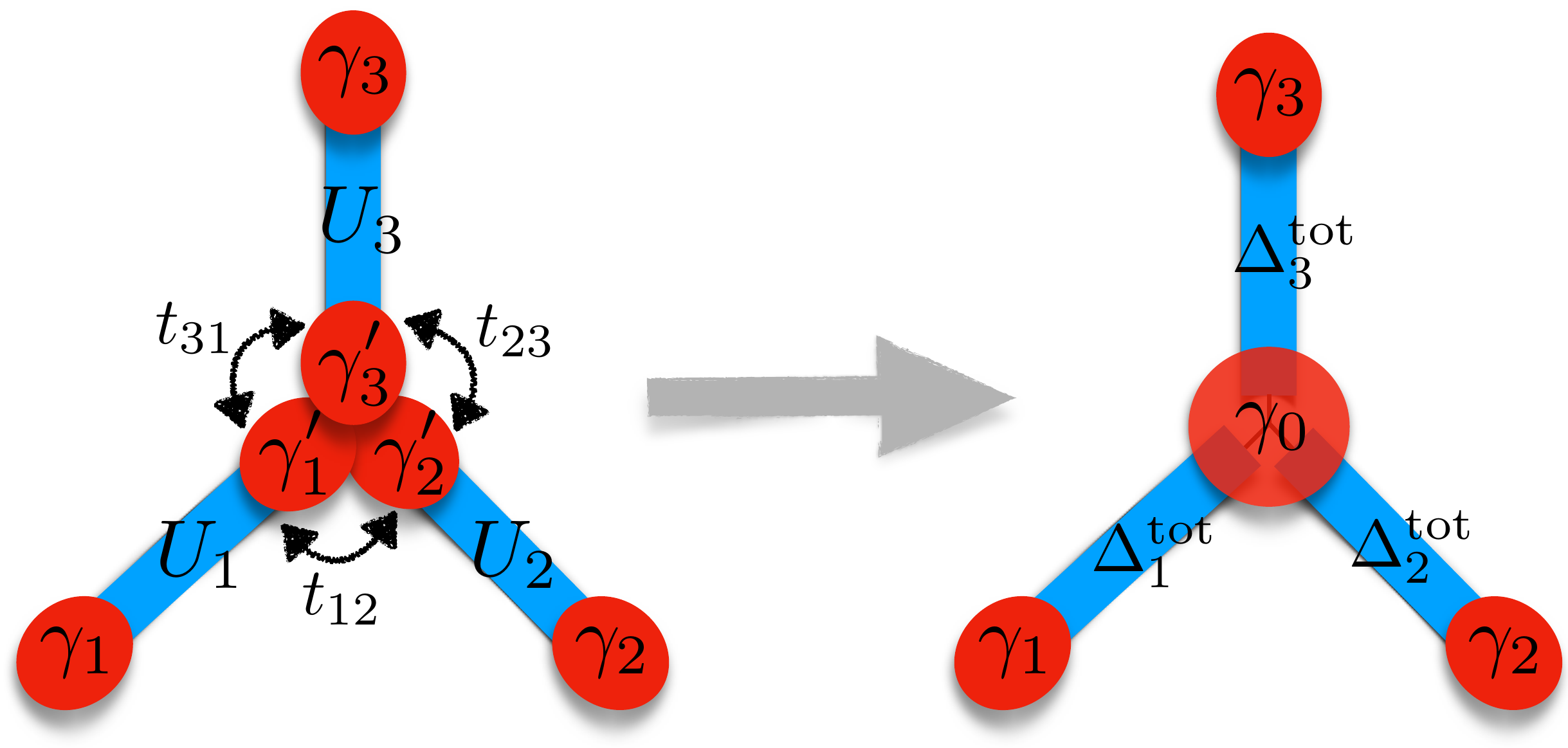}\
\caption{Sketch of the Y-junction setup and the resulting low-energy MBS subspace. Three topological wires hosting each MBS  at their ends ($\gamma_j,\gamma_j'$, with $j=1,2,3$) couple throughout the wires (with coupling strengths $U_i$) and between the wires (with coupling strengths $t_{ii+1}$) to give rise, in the middle, to an effective MBS $\gamma_0$. The coupling between the inner MBSs can be controlled, for example, by the fluxes through the three Josephson junctions to a bulk superconductor, which can contain a component from the cavity field that leads to the coupling  discussed in the text.}
\label{fig1} 
\end{figure}

Here we briefly review the Y-junction setup  proposed theoretically in many instances for implementing (non-abelian) braiding with MBS.  In Fig.~\ref{fig1} we show the Y-junction geometry that lends itself to braiding of MBS. All three wires consist of topological superconductors in the non-trivial phase, thus hosting MBSs at their edges.  The effective Hamiltonian describing such a Y-junction reads:
\begin{equation}
H_{\rm sys}=i\sum_{j=1}^3t_{jj+1}\gamma_j'\gamma_{j+1}'-i\sum_{j=1}^3U_j\gamma_j\gamma_j'\,,
\end{equation}  
where $\gamma_{j}'$ ($\gamma_{j}$) stand for the inner (outer) MBSs in wire $j=1,2,3$, $t_{jj+1}$ is the coupling between the inner MBSs of adjacent wires, $U_j$ are the coupling strengths between the MBSs within each wire (due to either direct wave-function overlap or due to Coulomb interaction effects). The tunnelings  $t_{jj+1}$ as well as the overlaps $U_i$ can be manipulated in various ways, depending on the actual implementation of the braiding. For example, they can be changed by affecting the fluxes on nearby Josephson junctions \cite{vanHeck2012}, by changing the orientation of an externally applied magnetic field, or by changing the electrostatic potential on the gates that define the wires  \cite{Alicea2011,Sau2011}. We define $E_M=\sqrt{\sum_jt^2_{jj+1}}$ as being the total energy associated with the coupling between the wires. 
Both the tunnelings and the MBS splittings can be affected in the presence of a cavity field ${\hat E}=\vec{E}_0(a^\dagger+a)$, with $\vec{E}_0$ and $a$ ($a^\dagger$) being the (vector) amplitude of the electric field  and the annihilation (creation) operator for photonic field in the cavity, respectively, so that:
\begin{align}
t_{jj+1}&\rightarrow t_{jj+1}+\delta t_{jj+1}(a^\dagger+a)\,,\\
U_j&\rightarrow U_j+\delta U_j(a^\dagger+a)\,,
\end{align}
with $\delta t_{jj+1}$ and $\delta U_{j}$ being the coupling strengths to the cavity that depend on the microscopic model. 

Next, we consider  $E_M\gg U_j$, and derive a new effective low energy Hamiltonian based on this approximation. First, we assume $U_j=0$, in which case one can define $\gamma_0=(1/\sqrt{3})(\gamma_1'+\gamma_2'+\gamma_3')$, such that one verify that this MBS, together with the other $3$ external ones drop out of the Hamiltonian; the other linear combinations of $\gamma_j'$s give rise to a two dimensional non-zero energy subspace by  $\pm(1/2) E_M$. In the case when the phases are such that not all of them are zero, we can disregard these states from the following discussion. Turning on the weak couplings between the inner and outer MBSs, one can now retain the leading order terms in $U_j/E_g$ only, so that we obtain the effective Hamiltonian:
\begin{align}
H_{Y}=\frac{i}{2}\sum_{j=1}^3\Delta_j\gamma_0\gamma_j\,,
\end{align}
where $\Delta_j=-(4t_{jj+1}/E_M)U_j$, in the absence of the cavity. When such a  coupling is considered, we need again to substitute $\Delta_j\rightarrow \Delta_j+\delta_j(a^\dagger+a)$, with $\delta_j$ depending on both $\delta t_{jj+1}$ and $\delta U_j$ (not shown). In fact, in order to account for the exponential coupling of the Majoranas, we assume in the following, as in Refs.~[\onlinecite{KnappPRX2016,SauCM2018}] that $\delta_i=\alpha_i\Delta_i$, with $\alpha_i$ the coupling strength between photon and the Majoranas. Physically, this means that far apart Majoranas cannot be affected by the cavity field. The low-energy effective  Hamiltonian becomes: 
\begin{align}
H_{Y}(t)=\frac{i}{2}\gamma_0[\vec{\Delta}+\vec{\delta}(a^\dagger+a)]\vec{\gamma}\,,
\end{align}
with $\vec{\Delta}=(\Delta_1,\Delta_2,\Delta_3)$,  $\vec{\delta}=(\delta_1,\delta_2,\delta_3)\equiv(\alpha_1\Delta_1,\alpha_2\Delta_2,\alpha_3\Delta_3)$, and $\vec{\gamma}=(\gamma_1,\gamma_2,\gamma_3)$. Note that in general  $\vec{\delta}\nparallel\vec{\Delta}$,  and thus there is dynamics triggered by the photonic field on the Majorana states. When considering braiding,  we need to consider the classical splittings time-dependent, or $\vec{\Delta}\rightarrow\vec{\Delta}(t)$.

One can define $\gamma_{\Sigma}=(1/E)\sum_{j=1}^3\Delta_j\gamma_j$, with $E=(1/2)\sqrt{\Delta_1^2+\Delta_2^2+\Delta_3^2}$, and two other linearly independent combinations that drop out of the Hamiltonian and spanning a two-dimensional zero energy, degenerate subspace. We can see that in the case when only one coupling $\Delta_j$ is non-zero, the MBSs that form that zero-energy subspace are precisely the MBSs that have $\Delta_j=0$. Next we introduce usual fermionic operators in terms of the MBSs ones $c_1=(\gamma_1-i\gamma_2)/2$ and $c_2=(\gamma_0-i\gamma_3)/2$, so that:
\begin{align}
\gamma_1=c_1+c_1^\dagger;\,\,\,\gamma_2=i(c_1-c_1^\dagger);\,\,\,\gamma_3=i(c_2-c_2^\dagger);\,\,\,\gamma_0=c_2+c_2^\dagger\,.
\end{align} 
This in turn allows us to write the Hamiltonian in the basis $\{|00\rangle, c_1^\dagger|00\rangle,c_2^\dagger|00\rangle,c_1^\dagger c_2^\dagger|00\rangle\}$\,, which gives the following Hamiltonian:
\begin{eqnarray}
H_Y=\frac{1}{2}\left(
\begin{array}{cccc}
\Delta_3 & i\Delta_1-\Delta_2 & 0 & 0\\
-i\Delta_1-\Delta_2 & -\Delta_3 & 0 & 0\\
0 & 0 & \Delta_3 & -i\Delta_1-\Delta_2\\
0 & 0 & i\Delta_1-\Delta_2 & -\Delta_3
\end{array}
\right)\,.
\end{eqnarray}
We see that the pair of states $\{|00\rangle,|11\rangle\}$ and $\{|01\rangle,|10\rangle\}$ do not couple with each other, which is a consequence of the conservation of the parity of the system
\begin{equation}
\tau_z=\gamma_0\gamma_1\gamma_2\gamma_3\,.
\end{equation} 
We can thus recast the above $4\times4$ Hamiltonian in terms of spin degrees of freedom (up to some spin rotations):
\begin{equation}
H_Y(t)=\frac{1}{2}[-\Delta_2\sigma_x+\tau_z\Delta_1\sigma_y+\Delta_3\sigma_z]\,,
\end{equation}
where $\sigma_{x,y,z}$ acts within the same parity states, while $\tau_{x,y,z}$ acts between different parities. In the presence of the cavity, the Hamiltonian for a given parity reads:
\begin{align}
H_Y^\tau(t)=\frac{1}{2}[\vec{\Delta}_\tau(t)+\vec{\delta}_\tau(a^\dagger+a)]\cdot\vec{\sigma}\,,
\end{align}
with $\vec{\Delta}_\tau(t)=[-\Delta_2(t),\tau\Delta_1(t),\Delta_3(t)]$, $\vec{\delta}_\tau(t)=[-\delta_2(t),\tau\delta_1(t),\delta_3(t)]$, and $\tau=\pm1$ being the eigenstates of $\tau_z$. Note that the resulting spectrum is the same for both parities. Written in terms of spin, this Hamiltonian corresponds to a time-dependent Rabi model, and in the following we focus on the dynamics. 

\section{Effective Majorana-Rabi model}

Here we describe the dynamics associated with the  effective MBS Rabi model for the two parities $\tau=\pm1$. The time-dependent Hamiltonian is $H_{Y}^\tau=H_{Y,0}^\tau(t)+H_{Y-ph}^\tau(t)+\omega_0 a^\dagger a$, with:
\begin{align}
H_{Y,0}^\tau(t)&=\frac{1}{2}\vec{\Delta}_\tau(t)\cdot\vec{\sigma}\,,\\
H_{Y-ph}^\tau&=\frac{1}{2}\vec{\delta}_\tau(t)\cdot\vec{\sigma}(a^\dagger+a)\,,
\end{align}
the last term being the MBS-photon coupling. As in the MT, it is instructive to work in spherical coordinates, and  define $\vec{\Delta}_\tau(t)=\Delta(t)\,\vec{n}^\tau_\Delta(t)$, and $\vec{\delta}_\tau(t)=\delta\,\vec{n}_\delta^\tau(t)$, with $\Delta(t)=\sqrt{\sum_{i}\Delta_i^2(t)}$,  $\delta(t)=\sqrt{\sum_{i}\delta_i^2}$, and $\vec{n}^\tau_{\alpha}=(\sin{\theta_\alpha}\cos{\phi_\alpha},\tau\sin{\theta_\alpha}\sin{\phi_\alpha},\cos{\theta_\alpha})$, where $\alpha=\Delta,\delta$ (nevertheless, in the following we drop the $\Delta$ index from the angles, while keeping the $\delta$ one). More specifically, we get:
\begin{align}
\delta(t)&=\Delta(t)\sqrt{[\alpha_1\sin\theta\cos\phi]^2+[\alpha_2\sin\theta\sin\phi]^2+[\alpha_3\cos\theta]^2}\,,\\
\cos{\theta_\delta}&=\frac{\alpha_3\cos\theta}{\sqrt{[\alpha_1\sin\theta\cos\phi]^2+[\alpha_2\sin\theta\sin\phi]^2+[\alpha_3\cos\theta]^2}}\,,\\
\tan{\phi_\delta}&=\frac{\alpha_1}{\alpha_2}\tan\phi\,,
\end{align}
so that in general $\theta(t)\neq\theta_\delta(t)$ as well as $\phi(t)\neq\phi_\delta(t)$. We perform a time-dependent unitary transformation $A_\tau(t)$ on the Y junction  Hamiltonian, where the columns of the matrix $A_\tau(t)$ are given by the instantaneous eigenstates of $\vec{\Delta}_{\tau}(t)\cdot\vec{\sigma}$:
\begin{align}
{\bf n}_\uparrow^\tau&=[\cos(\theta/2), e^{i\tau\phi}\sin{(\theta/2)}]\,,\\
{\bf n}_\downarrow^\tau&=[-\sin(\theta/2), e^{i\tau\phi}\cos{(\theta/2)}]\,,
\end{align}
The new transformed Hamiltonian becomes $\tilde{H}_{\rm Y}(t)=A_\tau^\dagger(t)H_{\rm Y}(t)A_\tau(t)-iA^\dagger_\tau(t)\partial_tA_\tau(t)$, or in more details it takes the form showed in the MT:
\begin{align}
\!\tilde{H}_{Y}^\tau(t)&=\frac{\Delta(t)}{2}\sigma_z+\tau\frac{\dot{\phi}}{2}(1+\cos{\theta}\sigma_z+\sin{\theta}\sigma_x)+\frac{\dot{\theta}}{2}\sigma_y+\frac{1}{2}A_\tau^\dagger(t)[\vec{\delta}_\tau(t)\cdot\vec{\sigma}]A_\tau(t) (a^\dagger+a)+\omega_0a^\dagger a\,.
\end{align}  
While the above expression is exact, next we focus only in the adiabatic limit with respect to the direct spin transitions, while we still allow for spin flips assisted by the cavity in the Hamiltonian. In such a case, $\dot{\phi},\dot{\theta}\ll\Delta(t)$ (at all times), and we can neglect the off-diagonal terms. Moreover, while not crucial to our description, we perform the rotating wave approximation to derive an effective MBS Jaynes-Cummings Hamiltonian, as showed in the MT:
\begin{align}
\!\!\tilde{H}^\tau_{\rm Y}(t)&\approx\frac{\Delta^{\rm eff}_\tau(t)}{2}\sigma_z+[\tilde{\delta}_\tau(t)a\sigma_++{\rm h. c. }]+\omega_0a^\dagger a\,,\\
\tilde{\delta}_\tau(t)&=\delta(t)\left[\cos{\theta_\delta}\sin{\theta}-\sin{\theta_\delta}\cos{\theta}\cos{(\phi-\phi_\delta)}+i\tau\sin{\theta_\delta}\sin{(\phi-\phi_\delta)}\right]/2\,,
\end{align}    
where $\Delta^{\rm eff}_\tau(t)=\Delta(t)+\tau\dot{\phi}\cos{\theta}$. The effective MBS splitting, $\Delta^{\rm eff}_\tau(t)$, contains a Berry phase contribution with opposite signs for opposite parities $\tau$, and that $\tilde{\delta}_-(t)=\tilde{\delta}^*_+(t)$, which also implies that  $|\tilde{\delta}_+(t)|=|\tilde{\delta}_-(t)|\equiv\tilde{\delta}(t)$ (independent of $\tau$).  In this work, we only focus on the dispersive regime,  quantified by the  condition $\tilde{\delta}_\tau(t)\ll|\Delta_\tau^{\rm eff}-\omega_0|$, and we can treat the coupling of the MBSs  to the cavity in time-dependent perturbation theory. To ease the calculation, we write
\begin{align}
\tilde{H}^\tau_{\rm Y}(t)=\tilde{H}_0(t)+V(t)\,,\\ 
V(t)=\tilde{\delta}_\tau(t)a\sigma^++\tilde{\delta}_\tau^*(t)a^\dagger\sigma^-\,,
\end{align}
and $\tilde{H}_0(t)$ the unperturbed part. It is instructive to perform a time-dependent Schrieffer-Wolff (unitary) transformation $U(t)=\exp{[S(t)]}$, with $S^\dagger(t)=-S(t)$ on the Hamiltonian $\tilde{H}_{\rm tot}(t)$, which pertains to a transformed one  $\bar{H}_{\rm tot}(t)=U^\dagger(t)\tilde{H}_{\rm tot}(t)U(t)-iU^\dagger(t)\partial_tU(t)$. We can now expand this Hamiltonian in powers of $V(t)$ and $S(t)$, and choose $S$ such that the off-diagonal contributions disappear. That is implemented by invoking 
\begin{align}
V(t)+[H_0(t),S(t)]-i\partial_t S(t)=0\,,
\end{align}  
which in turn allows us to write $S(t)=\tilde{e}_\tau(t)\sigma^+a-\tilde{e}_\tau^*(t)a^\dagger\sigma^-$. With such a form for the transformation, we can cast the resulting Hamiltonian into the following form:
\begin{align}
\bar{H}_{\rm tot}(t)\approx H_0(t)+\frac{1}{2}[S(t),V(t)]=\frac{\Delta^{\rm eff}_\tau(t)}{2}\sigma_z+\omega_0a^\dagger a+g_\tau(t)\left(a^\dagger a+\frac{1}{2}\right)\sigma_z\,,
\end{align}
where $g_{\tau}(t)={\mathcal Re}[\tilde{e}^*_\tau(t)\tilde{\delta}_\tau(t)]$, and where we neglected terms that are $\mathcal{O}(V^4)$. We are thus left with finding the function $e_\tau(t)$, which is found to  satisfy the following equation:
\begin{align}
[\omega_0-\Delta^{\rm eff}_\tau(t)]\tilde{e}_\tau(t)+\tilde{\delta}_{\tau}(t)+i\dot{\tilde{e}}_\tau(t)=0\,.
\end{align}
The general solution can be casted as:
\begin{align}
\tilde{e}_\tau(t)&=ie^{i\int^tdt'[\omega_0-\Delta^{\rm eff}_\tau(t')]}\int^tdt'\tilde{\delta}_\tau(t')e^{i\int^{t'}dt''[\omega_0-\Delta^{\rm eff}_\tau(t'')]}\,,
\end{align}
which in general cannot be fully solved analytically.  In this work we assume periodic Hamiltonians (as it is the case for the braiding discussed in this paper, and which is analyzed in the continuous precession limit). Moreover, as in the MT, we also assume $\Delta(t)=\Delta_0=const$, $\tilde{\delta}_\tau(t)=\tilde{\delta}_\tau(t+T)$,  
which allows us to write the Fourier decomposition:
\begin{align}
\delta_\tau(t)=\sum_{n}e^{in\Omega t}\delta^n_\tau\,,
\end{align}   
with $\tilde{\delta}_\tau^n$  being the $n$th Fourier components of $\tilde{\delta}_\tau(t)$. Utilizing the above expression, we find
\begin{align}
\tilde{e}_\tau(t)&=\sum_{n}\frac{\tilde{\delta}_\tau^{n}}{\Delta_0-\omega_0+(n+\tau\gamma_B/2\pi)\Omega}e^{in\Omega t}\,.
\end{align}
We mention  that in the limit  $\Omega\rightarrow0$, so that  we get
\begin{align}
\tilde{e}_\tau&=\frac{\tilde{\delta}_\tau}{\omega_0-\Delta_0}\,,
\end{align} 
which is the usual expression for the static case, independent on the parity and Berry phases. Utilizing the expression  for $\tilde{e}_\tau(t)$, we find:
\begin{align}
g_{\tau}(t)=\sum_{n,m}\frac{{\mathcal Re}[\tilde{\delta}_\tau^{n}\tilde{\delta}^{m*}_\tau e^{i(n-m)\Omega t}]}{\Delta_0-\omega_0+(n+\tau\gamma_B/2\pi)\Omega}\,.
\end{align}  
In order to fully  reveal the effect of the parity, it is instructive to write $\tilde{\delta}_\tau(t)=\tilde{\delta}_{r}(t)+i\tau\tilde{\delta}_i(t)$, with $\tilde{\delta}_{r(i)}(t)$ being the real (imaginary) parts of the coupling to the photons, so that $\tilde{\delta}_\tau^n=\tilde{\delta}_{r}^n+i\tau\tilde{\delta}^n_i$. That in turn leads to:
\begin{align}
g_{\tau}(t)&=\sum_{n,m}\frac{{\mathcal Re}[(\tilde{\delta}_{r}^n+i\tau\tilde{\delta}^n_i)(\tilde{\delta}_{r}^{m*}-i\tau\tilde{\delta}^{m*}_i)e^{i(n-m)\Omega t}]}{\Delta_0-\omega_0+(n+\tau\gamma_B/2\pi)\Omega}\nonumber\\
&=\sum_{n,m}\frac{{\mathcal Re}\left[\left(\tilde{\delta}_{r}^n\tilde{\delta}_{r}^{m*}+\tilde{\delta}_{i}^n\tilde{\delta}_{i}^{m*}-i\tau(\tilde{\delta}_{r}^n\tilde{\delta}_{i}^{m*}-\tilde{\delta}_{i}^{n}\tilde{\delta}_{r}^{m*})\right)e^{i(n-m)\Omega t}\right]}{\Delta_0-\omega_0+(n+\tau\gamma_B/2\pi)\Omega}\,.
\end{align}   
In this work, we are interested in the evolution over one period pertaining to this coupling, so we evaluate:
\begin{align}
g^\tau_{\gamma_B}=(1/T)\int_0^Tdtg_{\tau}(t)=\sum_{n}\frac{|\tilde{\delta}_{r}^n|^2+|\tilde{\delta}_{i}^n|^2+2\tau{\mathcal Im}[\tilde{\delta}_{r}^n\tilde{\delta}_{i}^{n*}]}{\Delta_0-\omega_0+(n+\tau\gamma_B/2\pi)\Omega}\,,
\end{align}
which is precisely the expression in the MT. We can see that in the static limit $\Omega=0$ the parity $\tau$ drops out of the evolution since $\tilde{\delta}_{r}^n\tilde{\delta}_{i}^{n*}\propto\delta_{n,0}$, and it is purely real. 


\section{Input-Output description and dissipation}

The read out of the Berry phase imprinted into the photonic field can be achieved  through dispersive read out of the cavity field in the input-output framework. For simplicity, we assume (like in the MT) that the cyclic braiding trajectory is performed in a continuous fashion.  Dissipation in MBS  systems in the static case, or during the braiding have been already studied by many authors (see, for example,  Refs.~\onlinecite{Pedrocchi2015a,Pedrocchi2015b,KnappPRX2016,Bauer2018,SauCM2018}), so here we only give a very short account of such effects. 

The coupling of the cavity to the external world (or stripline) that allows of the cavity to be probed reads:
\begin{align}
H_{c-b}&=-i\sum_{k,p=L,R}(f_{kp}a^\dagger b_{kp}-f_{kp}^*b^\dagger_{kp}a)\,,
\end{align}   
where $b_{kp}$ stand for the external line modes with wavevector $k$ and in output $p=L,R$ (assuming a two-sided cavity).  Following the standard procedures \cite{Dmytruk2015,kohlerPRL2018}, we can write  the equation of motion for the cavity field in the presence of the outside modes and the MBS in the spin language:
\begin{align}
\dot{a}(t)=-i\omega_{0}a(t)-\sum_p\frac{\kappa_p}{2}a(t)-\sum_p\sqrt{\kappa_p}b_{in,p}(t)-i(\vec{\delta}_\tau\cdot\vec{\sigma})(t)\,,
\end{align} 
where all operators are expressed in the interaction picture with respect to the full cavity and MBS Hamiltonian.  In the weak coupling limit, the spin dependence can be approximated as: 
\begin{align}
(\vec{\delta}_\tau\cdot\vec{\sigma})(t)\simeq(\vec{\delta}_\tau\cdot\vec{\sigma})^I(t)+ig\int_{-\infty}^tdt'[(a+a^\dagger)(\vec{\delta}_\tau\cdot\vec{\sigma})^I(t'),(\vec{\delta}_\tau\cdot\vec{\sigma})^I(t)]\,,
\end{align}  
where the index $I$ stands for the interaction picture with respect to the MBS system only (without the cavity, but considering various baths that give rise to relaxation and decoherence), and we assumed the initial time $t_0\rightarrow-\infty$. We can take the expectation value for the spin correlator as the errors we make are going to make will be of at least $|\vec{\delta}|^3$ (in the absence of the cavity, linear response):
\begin{align}
[(a+a^\dagger)(\vec{\delta}_\tau\cdot\vec{\sigma})^I(t'),(\vec{\delta}_\tau\cdot\vec{\sigma})^I(t)]&\simeq i\chi_{\tau}(t,t')[a(t')+a^\dagger(t')]\,,\\
\chi_{\tau}(t,t')&=-i\theta(t-t')\langle[(\vec{\delta}_\tau\cdot\vec{\sigma})^I(t'),(\vec{\delta}_\tau\cdot\vec{\sigma})^I(t)]\rangle_0\,,
\end{align}
with the last expression being the time-dependent susceptibility of the qubit in the absence of the cavity for a given parity $\tau$. For periodic driving, the susceptibility satisfies $\chi_\tau(t+T,t'+T)=\chi_\tau(t,t')$, and one defines $\chi_\tau(t,t-\tau)$, with $\tau=t-t'$, so that:
\begin{equation}
\chi_{\tau}(t,t-\tau)=\sum_p\int \frac{d\omega}{2\pi}e^{-ip\Omega t-i\omega\tau}\chi_{\tau}^{(p)}(\omega)\,,
\end{equation}
which in turn gives the following frequency dependence of the spin:
\begin{equation}
(\vec{\delta}_\tau\cdot\vec{\sigma})(\omega)=(\vec{\delta}_\tau\cdot\vec{\sigma})^I(\omega)-\sum_p\chi_{\tau}^{(p)}(\omega-p\Omega)[a(\omega-k\Omega)+a^\dagger(\omega-p\Omega)]\,.
\end{equation}
Substituting this expression into frequency domain formula, and neglecting the fluctuating term $(\vec{\delta}_\tau\cdot\vec{\sigma})^I(\omega)$ in the limit of large number pf photons, we obtain \cite{kohlerPRL2018}:
\begin{align}
i(\omega_0-\omega-i\kappa_{tot}/2) a(\omega)=-\sum_p\sqrt{\kappa_p}b_{in,p}(\omega)+i\sum_k\chi_{\tau}^{(k)}(\omega-k\Omega)[a(\omega-k\Omega)+a^\dagger(\omega-k\Omega)]\,.
\end{align}
We see that the cavity photons at frequency $\omega$ are mixed with those are multiple of the qubit driving frequency $\Omega$. However, close to resonance ($\omega\approx\omega_0$), in the weak damping limit, and assuming a good cavity  $\Omega\gg\kappa_{tot}$, we can safely keep in the above expression only the terms with $k=0$, as well as neglecting the $a^\dagger(\omega-k\Omega)$ term (again, off-resonant). In this case, we eventually obtain:   
\begin{equation}
a(\omega)=i\sum_p\frac{\sqrt{\kappa_p}}{\omega_0-\omega-i\kappa_{tot}/2-\chi_{\tau}^{(0)}(\omega)}b_{in,p}(\omega)\,.
\end{equation}
This can be  combined with the input-output relations  $b_{in,p}(\omega)+b_{out,p}(\omega)=\sqrt{\kappa_p}a(\omega)$ to finally give
\begin{equation} 
\langle b_{out,2}(\omega)\rangle=\frac{-i\sqrt{\kappa_{1}\kappa_2}}{\omega-\omega_0+i\frac{\kappa_{tot}}{2}-\chi^{(0)}_{\tau}(\omega)}\langle b_{in,1}(\omega)\rangle\,,
\end{equation}
where we assumed a flux of probing photons  is sent into the cavity from one side, say left, and we look at the photons transmitted on the right side. The transmission is then defined as $t(\omega,\Omega)=\langle b_{out,2}(\omega)\rangle/\langle b_{in,1}(\omega)\rangle$, which is  the expression showed in the MT, but with the $g_{\gamma_B}^\tau(\omega,\Omega)\rightarrow-\chi_{\tau}(\omega,\Omega)$. Next we show that these two are the same in the adiabatic limit.    

While the full description should be made in terms of the stationary (but time-dependent) density matrix, here we address this problem first without any dissipation at all (at the level of the wavefunction).  The parity-dependent Schrodinger equation reads:
\begin{equation}
i\frac{\partial |\Psi_\tau(t)\rangle}{\partial t}=H_0^\tau(t)|\Psi_\tau(t)\rangle\,,
\end{equation}
which, when switching to the rotating frame, allows us to write $|\Psi_\tau(t)\rangle=A(t)|\psi_\tau(t)\rangle$, where $|\psi_\tau(t)\rangle$ is given by
\begin{equation}
|\psi_\tau(t)\rangle=\underbrace{e^{-\frac{i}{2}\sigma_z(\Delta_0+\tau\gamma_B/T)t}}_{U_\tau(t)}|\psi_\tau(0)\rangle\,,
\end{equation} 
where $|\psi_\tau(0)\rangle$ is the initial state. Putting the entire evolution together, we get:
\begin{equation}
U_\tau(t,t')=\sum_{\sigma}e^{-\frac{i}{2}\sigma(\Delta_0+\tau\gamma_B/T)(t-t')}A(t)|\psi_\tau^\sigma\rangle\langle\psi_\tau^\sigma|A^\dagger(t')\,,
\end{equation}
which we can use to deal with the interaction picture operators in the expression for the susceptibility. We  assume that the system was initialized in the instantaneous ground state, and  that it stays in the ground state during the evolution. Then, the density matrix of the qubit is given by $\rho(t)=A(t)|\psi_{\tau}\rangle\langle\psi_{\tau}|A^\dagger(t)$, with $|\psi_\tau\rangle$ being the instantaneous ground state function for parity $\tau$. We can the write for the susceptibility:
\begin{align}
\chi^{(p)}_{\tau}(\omega)&=-\int_{0}^\infty ds\int_0^{2\pi} \frac{d(\Omega t)}{2\pi} e^{i\omega s+ip\Omega t}\chi_{\tau}(t,t-s)\nonumber\\
&=i\int_{0}^\infty ds\int_0^{2\pi} \frac{d(\Omega t)}{2\pi} e^{i\omega s+ip\Omega t}\langle\psi_\tau|[(\vec{\delta}\cdot\vec{\sigma})^I(t),(\vec{\delta}\cdot\vec{\sigma})^I(t-s)]|\psi_\tau\rangle\,.
\end{align}
Let us now evaluate the correlator:
\begin{align}
\langle\psi_\tau|[(\vec{\delta}\cdot\vec{\sigma})^I(t),(\vec{\delta}\cdot\vec{\sigma})^I(t')]|\psi_\tau\rangle&=e^{-i(\Delta_0+\tau\gamma_B/T)(t-t')}\langle\psi_\tau|A(t)(\vec{\delta}\cdot\vec{\sigma})A^\dagger(t)|\bar{\psi}_{\tau}\rangle\langle\bar{\psi}_{\tau}|A(t')(\vec{\delta}\cdot\vec{\sigma})A^\dagger(t')|\psi_{\tau}\rangle\nonumber\\
&-e^{i(\Delta_0+\tau\gamma_B/T)(t-t')}\langle\psi_\tau|A(t')(\vec{\delta}\cdot\vec{\sigma})A^\dagger(t')|\bar{\psi}_{\tau}\rangle\langle\bar{\psi}_{\tau}|A(t)(\vec{\delta}\cdot\vec{\sigma})A^\dagger(t)|\psi_{\tau}\rangle\nonumber\\
&=e^{-i(\Delta_0+\tau\gamma_B/T)(t-t')}\tilde{\delta}_\tau(t)\tilde{\delta}^*_\tau(t')-e^{i(\Delta_0+\tau\gamma_B/T)(t-t')}\tilde{\delta}_\tau(t')\tilde{\delta}^*_\tau(t)\nonumber\\
&=\sum_{k,k'}\left[e^{-i(\Delta_0+\tau\gamma_B/T+k'\Omega)s}-e^{i(\Delta_0+\tau\gamma_B/T-k\Omega)s}\right]e^{i(k+k')\Omega t}\tilde{\delta}^{k}_{\tau}(\tilde{\delta}^{k'}_{\tau})^*\,,
\end{align}
where $s=t-t'$, and $\tilde{\delta}_\tau(t)$ is the coefficient of $\sigma_\tau$ in the expression $A(t)(\vec{\delta}\cdot\vec{\sigma})A^\dagger(t)$, which precisely corresponds to the RWA terms in the MT [as well as fullfilling $\tilde{\delta}_\tau(t)=\tilde{\delta}_{\bar{\tau}}^*(t)$]. Moreover, we defined:
\begin{align}
\tilde{\delta}^{k}_{\tau}=\frac{1}{T}\int_{0}^Tdt e^{ik\Omega t}\tilde{\delta}_\tau(t)\,,
\end{align}
so that we get for the susceptibility:
\begin{align}
\chi^{(p)}_\tau(\omega)&=\sum_{k,k'}\left[\frac{1}{\omega-\Delta_0-\tau\gamma_B/T-k'\Omega+i\eta}-\frac{1}{\omega+\Delta_0+\tau\gamma_B/T-k\Omega+i\eta}\right]\tilde{\delta}^{k}_{\tau}(\tilde{\delta}^{k'}_{\tau})^*\delta_{k-k',p}\,.
\end{align}
or, for $p=0$, as we need in our description
\begin{align}
\chi^{(0)}_\tau(\omega)&=\sum_{k}\left[\frac{1}{\omega-\Delta_0-\tau\gamma_B/T-k\Omega+i\eta}-\frac{1}{\omega+\Delta_0-\tau\gamma_B/T+k\Omega+i\eta}\right]|\tilde{\delta}^{k}_{\tau}|^2\nonumber\\
&\approx\sum_{k}\frac{|\tilde{\delta}^{k}_{\tau}|^2}{\omega-\Delta_0-\Omega(k+\tau\gamma_B/2\pi)+i\eta}\equiv-g^\tau_{\gamma_B}(\omega+i\eta,\Omega)\,,
\end{align}
in Eq.~10 in the MT.  By taking into account a Markovian bath and a resulting finite population of the excited levels,  we can extend the description to a fully open system, in which case (in the weak coupling to the environment limit), we can write \cite{kohlerPRL2018}:
\begin{align}
\chi^{(p)}_\tau(\omega)&=\sum_{s=\pm, k'}\frac{(p_s-p_{\bar{s}})(\tilde{\delta}_\tau^{k'-p})^*\tilde{\delta}_\tau^{k'}}{\omega+s(\Delta_0+\tau\gamma_B/T)+k'\Omega+i\Gamma}\,,
\end{align}
where $p_{s}$ are the populations of the instantaneous levels $s$, and $\Gamma$ is their broadening (assumed constant, for simplicity).
